# Probing the Structure and *in Silico* Stability of Cargo Loaded DNA Icosahedron using MD Simulations

Himanshu Joshi[1], Dhiraj Bhatia[2], Yamuna Krishnan[3,4] and Prabal K. Maiti[1*]

1. *Centre for Condensed Matter Theory, Department of Physics, Indian Institute of Science, Bangalore 560012, India*

2. *Institut Curie, PSL Research University, Chemical Biology of Membranes and Therapeutic Delivery unit, INSERM, U 1143, CNRS, UMR 3666, 26 rue d'Ulm, 75248 Paris Cedex 05, France*

3. *Department of Chemistry, The University of Chicago, Chicago, Illinois 60637, USA*

4. *Grossman Institute of Neuroscience, Quantitative Biology and Human Behavior, The University of Chicago, Chicago, Illinois 60637, USA*

\* To whom correspondence should be addressed. Tel: +091-80-22932865
Fax: +91-80-23602602; Email: maiti@physics.iisc.ernet.in

**Abstract**

Platonic solids such as polyhedra based on DNA have been deployed for multifarious applications such as RNAi delivery, biological targeting and bioimaging. All of these applications hinge on the capability of DNA polyhedra for molecular display with high spatial precision. Therefore high resolution structural models of such polyhedra are critical to widen their applications in both materials and biology. Here, we present an atomistic model of a well-characterized DNA icosahedron, with demonstrated versatile functionalities in biological systems. We study the structure and dynamics of this DNA icosahedron using fully atomistic molecular dynamics simulation in explicit water and ions. The major modes of internal motion have been identified using principal component analysis. We provide a quantitative estimate of the radius of gyration ($R_g$), solvent accessible surface area (SASA) and volume of the icosahedron which is essential to estimate its maximal cargo carrying capacity. Importantly, our simulation of gold nanoparticles (AuNP) encapsulated within DNA icosahedra revealed enhanced stability of the AuNP loaded DNA icosahedra compared to empty icosahedra. This is consistent with experimental results that show high yields of cargo-encapsulated DNA icosahedra that have led to its diverse applications for precision targeting. These studies reveal that the stabilizing interactions between the cargo and the DNA scaffold powerfully positions DNA polyhedra as targetable nanocapsules for payload delivery. These insights can be exploited for precise molecular display for diverse biological applications.

**Introduction**

Structural DNA nanotechnology utilizes the key properties of DNA such as its persistence length and sequence specific recognition via Watson-Crick base pairing to create various architectures on the nanoscale.[1] One of the important classes of architectures fabricated by structural DNA nanotechnology is regular three-dimensional platonic solids or polyhedra. There are four different strategies to construct DNA polyhedra namely one pot assembly,[2] modular self-assembly,[3] hierarchical assembly[4] and the origami based approach.[5] DNA polyhedra afford unique advantages over other kinds of DNA scaffolds, which predispose the former for biological applications. First is their ability to encapsulate various nanoscale entities such as drugs, imaging agents or nanoparticles within their internal void for uses *in vivo*. The other equally exciting but thus far under explored property is their capability to act as molecular pegboards for site specific display of various ligand like small molecules, bio-tags like peptides, siRNAs, etc.[6, 7] DNA polyhedra have the capacity to encapsulate water soluble polymers such as FITC Dextran and inorganics such as gold nanoparticles (AuNPs) or quantum dots from solution.[3, 8-10] Further, cargo loaded DNA icosahedra have been explored for targeted, functional bioimaging in living cells and organisms.

The composition and morphology of DNA-based polyhedra have been verified at both bulk and single molecule levels using agarose gel electrophoresis (AGE), fluorescence resonance energy transfer (FRET), atomic force microscopy (AFM) and transmission electron microscopy (TEM) which provide coarse grained structural information.[11] While such studies confirm sequence connectivity and overall morphology, information on the structure and dynamics of the DNA polyhedron at single nucleobase resolution remains inaccessible.[12] All atom MD simulations of such DNA polyhedra could be extremely useful in providing this information at single nucleotide resolution. A simulation study with atomistic model would address factors such as tensegrity, structural dynamics and conformational flexibility of the relevant polyhedron that could aid better design strategies to assemble polyhedra in high yields. This information is critical in order to explore the full potential of such synthetic, payload-containing polyhedra to also act as a scaffold for precise molecular display. Substantial advances in computing capability and molecular force fields in past decade position atomistic simulation as a persuasive method to understand DNA based nanostructures.[13, 14] Simulation techniques have been applied to study various DNA nanostructures such as cross-over junctions[15, 16] truncated octahedron,[17, 18] origami structures[19, 20] and DNA nanotubes.[21, 22] Given the extensive structural and bio-functional experimental validation of an icosahedral DNA nanocapsule, we chose to explore its structure and dynamics at an atomistic level. 200 ns long fully atomistic MD simulations have been carried out with explicit solvent and ions to demonstrate structural stability of DNA architectures *in silico*. We follow the trajectories of the atoms to probe the properties of DNA icosahedron in solution. Further, for the first time, we explore a new level of complexity in DNA nanostructures by modeling a DNA icosahedron containing a cargo of gold nanoparticles (AuNPs) within its internal void. Our modeling accurately maps the position and orientation of different biological residues or tags e.g. folic acid, conjugated to specific nucleotide positions on the icosahedron. This comprehensive atomistic map of the DNA icosahedron now enables researchers to appropriately choose specific positions on this icosahedron to conjugate and display diverse molecular tags for various biological applications. Further, simulations on cargo loaded DNA icosahedra provide insight on the dynamics of the DNA scaffold that provides information's not only on polyhedra-cargo interactions, but also interaction of a tag-displaying, cargo-loaded icosahedron with its biological target. This could help better design of polyhedra-cargo pairs for a variety of application. We first describe the building methodology and simulation protocol followed by the various analyses schemes to quantify the structural and dynamical evolution of the simulated structures. Finally, we discuss the implications of our findings in the broader context of DNA nanotechnology. To the best of our knowledge, this is a first of its own kind of study on the experimentally characterized structure of the largest and most complex DNA based platonic structure using atomistic MD simulations to understand the in-solution behavior.

### *De novo* design: Building protocol

We have followed the experimental framework of Bhatia *et al.* for the design and sequences of the nucleotides comprising the DNA icosahedron.[3, 23] Eight cyclic single strands of DNA self-assemble to form an icosahedron with its 30 edges composed of double stranded DNA helices. Figure 1 shows a schematic representation of the structure and assembly strategy used to build the DNA icosahedron. The atomistic model of icosahedral DNA was constructed using the Nucleic acid builder (NAB),[24] a programming environment to construct nucleic acid structures with non-canonical topologies and the xLEaP module of the AMBER14 suite of programs.[25] First, we created the skeleton of an ideal convex icosahedron with 12 vertices in a virtual pdb file format. The coordinates of the vertices of the ideal icosahedron with the edge length "2" are given as $(0, \pm\phi, \pm1), (\pm\phi, \pm1, 0)$ and $(\pm\phi, 0, \pm1)$

$$\text{where } \phi = \frac{1+\sqrt{5}}{2}$$

In our model, the length of each arm is 12.0 nm, slightly longer (3.2 nm) than 26 bp. (see supplementary information (SI) Figure S1 for the details). Each vertex is a five-way junction (5WJ) with equilateral triangular facets, that serves as a guide to accurately align 26 bp long duplex DNA segments bearing 2 unpaired nucleotides on three duplexes at each vertex, (labeled V3-V5, L3-L5 and U3-U5 in the experiments).[3] The alignment of dsDNA duplex domains in an icosahedral geometry was performed using the rigid body matrix transformations functionality in NAB. Next, in order to obtain suitable phosphodiester bond lengths between the backbones of the adjacent duplex edges that converged at a given vertex, we perform a series of rotations to the individual dsDNA domains. Upon obtaining a feasible bond length, a phosphodiester bond between the abutting strands was created in xLEaP in accordance with the experimentally designed connectivity of the DNA icosahedron. The 2 unpaired bases (overhangs) at each of the three edges of all the 5WJ vertex of the icosahedron provide additional structural flexibility. These linkers play a significant role providing stability to the 5WJ motifs. Recently, Iacovelli *et al.* have shown using a computational study on truncated DNA octahedron that single stranded linkers provide additional flexibility to the DNA polyhedron as compared to the double stranded (dS) linker.[26] The junctions with different symmetric axes are explicitly shown in the Figure S2 in SI.

To study the host cargo interaction, we built face centered cubic (FCC) crystalline faceted gold nanoparticles (AuNPs) and placed them inside the DNA icosahedron. Materials studio 6.0[27] was used to obtain the desired facet [111] and [100] of the AuNPs. Each AuNP has 4455 atoms and we could maximally pack 21 such AuNPs inside a single DNA icosahedron. The DNA icosahedron with and without encapsulated AuNPs, were loaded in xLEaP module of AMBER molecular dynamics package and solvated with a truncated octahedral box of water using TIP3P water model.[28] Here, we ensured minimum 15 Å aqueous shell around the atoms of the DNA molecules. The system was charge neutralized by adding $Na^+$ ions around the duplex domains comprising the DNA icosahedron (see Figure S3 in SI for the ion placements). xLEaP module constructs a Coulombic potential on a grid of 1 Å resolution and then places ions one at a time at the lowest electrostatic potential near the phosphate groups of the DNA backbone. Additionally, 24 $Mg^{2+}$ ions and 48 $Cl^-$ ions were added to further stabilize the 5WJs. Figure 2(a) shows the NAB built atomistic model of icosahedral DNA and Figure 2(b) represents the solvated structure of AuNP encapsulated DNA icosahedron ($I_{AuNP}$) with ions. Table 1 summarizes the details of all the simulated systems. Considering the number of atoms and simulation time scale, we believe these simulations are the largest among all the reported MD studies on DNA nanostructures.

**Simulation Methodology**

All atom molecular dynamics simulations have been performed using AMBER14 molecular dynamics package.[25] The bonded and non-bonded description of the interactions between the various atoms has been described using the AMBER14 force fields which include the ff99[29, 30] force field parameters with parmbsc0[31] refinement for nucleic acids. This non-polarizable and additive force field utilizes the charge scheme given by Cornell el al.[29] We have incorporated the recently optimized Joung-Cheatham (JC) parameters[32] for $Na^+$, $Cl^-$ and Li *et al.* parameters[33] for $Mg^{2+}$ to account for the non-bonded interactions of these metal ions. These force fields are known to describe the structure and dynamics of nucleic acids accurately.[14] It would be interesting to study the effect of recently developed ion parameters by Yoo and Aksimentiev on the structure and stability of the DNA nanostructures.[34, 35] For AuNPs, we use the Lennard-Jones (LJ) parameters by Heinz *et al.*[36] Initially, we perform a series of energy minimization steps to eliminate any bad contacts in the initially built structures. During the minimization, DNA and AuNPs are restrained with harmonic force constants reducing from 500 kcal/mol/Å$^2$ to 0 through several minimization steps. Each minimization step involves 1000 steps of steepest descent followed by 2000 steps of conjugate gradient method. After the energy minimization, the system is slowly heated up to 300 K in 40 ps MD using 1 fs integration time step, while restraining the solute with 20 kcal/mol/Å$^2$ harmonic force constant. After this, we perform 5 ns NPT equilibration of the structures with no harmonic restraints. This process leads to correct solvent densities i.e. close to experimental values. Finally, 200 ns NPT production simulations are performed at 300 K and 1 atm pressure with 2 fs integration time step. We have implemented periodic boundary condition across the system using a truncated octahedral cell. We use Particle Mesh Ewald (PME) techniques integrated with AMBER package to account for the long range part of the electrostatic interactions.[37, 38] During the dynamics, all the bonds involving hydrogen are restrained using the SHAKE algorithm.[39, 40] Langevin thermostat with collision frequency of 1 ps$^{-1}$ is used to maintain the constant temperature while the pressure is controlled by anisotropic Monte-Carlo barostat.[41] To avoid known synchronization artifacts,[42] we used a different seed for the random number after each nano-second MD simulation. The translational and rotational motion of the center of mass (COM) were removed after 1000 MD steps. The accelerated GPU version of PMEMD[43, 44] was performed on Nvidia K40 series cards. A similar protocol was successfully used in our previous studies on other DNA nanostructures.[45-47] We have employed CPPTRAJ[48] functionality of AMBERTOOLs[25] to perform various analyses on the equilibrium MD simulation trajectories. The definition, terminology and procedure to calculate DNA geometrical parameters are similar to the 3DNA software.[49] The images and graphics of the structures shown here were generated using the software packages VMD[50] and PyMOL.[51]

**Results and Discussion**

**Structural Deviation and Atomic Fluctuation: RMSD and RMSF**

The time evolution of the Root-Mean-Square-Deviation (RMSD) of the DNA icosahedron is shown in Figure 3(a). We compute the RMSD of the structures with respect to the built energy minimized structures as well as with respect to the time-averaged structure (equivalent of solution structure) during the simulations. This reveals that the RMSD values converge in all cases confirming that the structures have attained their equilibrium configuration within 205 ns long simulation. The RMSD with respect to their initial minimized structure saturates to a value of 1.8 nm and 1.1 nm for $I_{empty}$ and $I_{AuNP}$ respectively. The RMSD with respect to the average simulation structure settles to ~0.7 nm and 0.3 nm for $I_{empty}$ and $I_{AuNP}$ respectively after 205 ns MD. The lower values of RMSD for the $I_{AuNP}$ are also reflected in the instantaneous snapshots of the structures shown in Figure 4 and Figure S4 in SI. Snapshots at various time instants of the simulation show that the $I_{AuNP}$ undergoes less structural deformation with respect to the built energy minimized structure during the course of MD simulations. Figure 3(b) shows Root-Mean-Square-Fluctuation (RMSF) in the atomic positions with respect to the individual edges of the icosahedron along with the standard deviation. The RMSF values are averaged over all atoms of the edges and also over entire MD simulation trajectories. The RMSF per edge for $I_{empty}$ varies from 5.2 Å to 11.6 Å whereas for $I_{AuNP}$, it varies from 3.6 Å to 6.5 Å. To closely follow the fluctuation in the atomic positions, we also calculated per-atom and per-residue RMSF (Figure S5 in SI). This analysis confirms that the AuNP encapsulation within the icosahedron reduces the flexibility of the edges of the icosahedron, providing structural stability. Owing to the conformational flexibility of the dsDNA edges of the DNA icosahedron, the icosahedral DNA nanocage shrinks during the initial course of the MD simulation. During this compaction process, some of the nucleotides of the edges of the DNA icosahedron experience steric hindrances due the nearby atoms of crystalline AuNPs. This leads to the decrease in the structural fluctuations of the respective dSDNA edge of the $I_{AuNP}$. This analysis explains the higher yields of $I_{AuNP}$ compared to $I_{empty}$ in the experiments. We have calculated the average helical parameters of the edges of icosahedron to quantify and compare the deviation in helical structure of DNA with respect to the reference B-DNA structure. We found that, apart from the fluctuations at the vertices, the overall B-DNA geometry of DNA is preserved during the course of simulation. Further, the helical parameters are better maintained in $I_{AuNP}$ as compared to $I_{empty}$ which mimics the higher stability of the former. The detail analysis is presented in Table S1 and Table S2 of the appendix S9 in SI.

**Structure and Dynamics of the Encapsulated AuNPs**

To study the stability of $I_{AuNP}$, we have used faceted AuNP loaded icosahedron. Faceted nanoparticles possesses larger surface to volume ratio and hence the enhanced interaction with the surroundings.[52] We created energetically most favorable facets in the FCC Au lattice using the material studio software as shown in the Figure 5(a). Each AuNP has 8 [111] and 6 [100] surfaces. 21 such faceted AuNPs are closely packed inside the icosahedral DNA using a spatial transformation in NAB. Figure 5(b) compares the radial distribution function (RDF) of AuNPs prior to and after the 200 ns MD simulation. Both the curves almost overlap, suggesting that there is no structural change in the crystalline AuNPs in the course of MD simulation. The peaks in the RDF correspond to the distance of first nearest neighbor shells, second nearest neighbor shells and so on. These peaks are very similar to the RDF peaks in the bare crystalline AuNPs assembly investigated recently.[53] The structure of AuNP remains intact during the entire course of simulation. The self-interaction energy of AuNPs, which comes out to be ~175 kcal/mol,[53] is much stronger compared to the interaction energy with the surrounding DNA. Thus the strong binding energy of AuNPs governs the intact structure of AuNPs confined inside the DNA icosahedron. Radius of gyration of all the AuNPs inside the icosahedral DNA is shown on the top right side of the Figure 5 (b). This is uniformly preserved at 54.4 Å throughout the simulation. The simulation snapshots (Figure 4 and Figure S4 in SI) confirm increased stability of the encapsulated gold nanoparticles, indicating that AuNP structure is unaffected due to confinement within the icosahedral DNA shell.

**Evolution of the End to End Distance**

From the atomistic MD simulation, one can extract important structural and dynamical information about the DNA icosahedron to inform future designs using this polyhedron as a scaffold for molecular display. The icosahedron, also represented as [3 5], is the most complex among the family of five platonic solids with six diagonals. The theoretical value of the diagonal distance for an ideal convex icosahedron with ~12 nm edge length is ~20 nm. To quantify the structural changes, we also measure the distances of two diagonals (out of six diagonals) of icosahedral DNA. Here to define the diagonal distance, we consider all of the five terminal base pairs at corresponding vertex i.e. 5WJ, find the center of geometry and measure the distance between the geometrical centers of atoms defined in similar way on opposite ends. The scheme of the calculation is shown in Figure 6(a), D1 and D2 are the two diagonal distances between the vertices highlighted in red and blue. The time evolution of the diagonal lengths during 205 ns long simulation is shown in Figure 6(b). For $I_{empty}$, D1 decreases to ~17.2 nm within few ns of simulation and oscillates around 18.0 nm towards the end of 205 ns simulation. On the other hand, D2 increases initially, fluctuates around 20 nm and later reduces to ~18.5 nm. In case of $I_{AuNP}$, D1 and D2 follow a trend similar to $I_{empty}$ but the RMS fluctuations are much less. After 205 ns simulation, D1 and D2 equilibrate to ~ 17.5 nm and ~ 18.7 nm respectively for both the structures. These diagonal distances seem to be most favored for the structure of the icosahedral DNA assembly and are consistent with the experimentally reported values measured using transmission electron microscope (TEM).[3]

**Major Modes of Dynamics using Principal Component Analysis**

To account for the major modes of motion and compare the positional fluctuation in the dynamics of the simulated structures quantitatively, we have implemented principal component analysis (PCA) technique.[54-56] PCA is designed to figure out the dominant internal motion among the various components of the systems. This robust technique is widely used to study the convergence of MD simulation in phase space.[57] We deployed it to filter out any redundant modes and extract the main modes of motion to quantify structural fluctuations in the DNA icosahedron. In the process, we formulate the covariance matrix from the Cartesian coordinates of 1632 phosphorus atoms comprising the DNA icosahedron during the course of the MD simulation. The coordinates have been fitted to the average structure to remove overall rotational and translational motion. We then diagonalized the covariance matrix and calculated 4896 eigenvalues and eigenvectors. The top principal components which are the low frequency modes correspond to the bending and twisting motion of the double helical edges hinged around the junction. Figure 7(a) represents the first principal component *i.e* the direction along the largest structural deviation in the DNA icosahedron during MD simulation. The corresponding vector is shown using an arrow to indicate the average position of P atom in the normal mode wizard tool in VMD. The eigenvalues statistics, (see Figure S6 in SI) reveals that out of a total of 4896 modes, the first principal component accounts for 38% and 47%of the total motion in the $I_{empty}$ and $I_{AuNP}$ respectively. Further, the first 5 principal components cover ~ 65% of the total motion of both the structures. The trajectories of the simulation are projected along first 5 principal components and histogram of the projection is shown in the Figure 7(b) and 7(c). This indicates how far the structure has progressed along the corresponding mode in the new coordinate system. The empty icosahedron $I_{empty}$ maintains a higher projection and larger spread along the various principal components than $I_{AuNP}$ which further confirms reduced internal motion in the latter.

**Surface and Volume Analysis**

The DNA icosahedron has been used to encapsulate and deliver a whole range of molecular cargo in cells and *in vivo*. Thus its molecular surface area and volume are very important in order to estimate its true loading capacity as well as to understand cargo-capsule interaction with its surrounding entities. To account for the variation in overall size and volume of the DNA icosahedron during the simulation, we have calculated its radius of gyration ($R_g$) using the following definition for a molecule with n number of atoms.

$$R_g = \sqrt{\frac{1}{N}\sum_{i=1}^{N}(r_k - r_{mean})^2}$$

Where, $r_{mean}$ is the mean position of all the n atoms at that instant of time.

Figure S7(a) shows the time evolution of $R_g$ which is initially at 95 Å and reduces during the simulation to ~ 87 Å and 90 Å for $I_{empty}$ and $I_{AuNP}$ respectively. The analysis reveals a minor compaction with respect to the built structure that probably arises from breathing modes due to flexibility in the DNA icosahedra. Successful formation of well-defined

DNA nanostructures requires a degree of flexibility at the junctions and rigidity of the duplex DNA domains. Though both the architectures follow the same trend, $I_{AuNP}$ shows less compaction, attributed to reduced architectural flexibility arising from the encapsulated AuNPs. $R_g$ is related to the volume enclosed within the DNA icosahedron. The volume associated with this radius is ~ 3000 nm$^3$ which is in good agreement with experimental determinations[3] and the computed volume of our model using Voronoi diagram based geometrical methods to compute the void volume.[58] The encapsulated volume of the simulated structures is (~2500 nm$^3$) consistent with analytical and experimental observations. To investigate the evolution of the structure, we have also calculated the solvent accessible surface area (SASA) of DNA icosahedron using the linear combinations of pairwise overlap (LCPO) method.[59] SASA is plotted as a function of simulation time in Figure S7(b) and shows an initial value of 2729 nm$^2$, that deceases to 2686 nm$^2$ and 2696 nm$^2$ for the $I_{empty}$ and $I_{AuNP}$ respectively. The accessible solvent area initially dips in the $I_{AuNP}$ and thereafter stays fairly uniform with time. In the case of $I_{empty}$, SASA decreases initially, stabilizing at ~2690 nm$^2$ during the final 25 ns of simulation. These analyses suggest that the DNA icosahedron is closely follow ideal geometry of icosahedrons during the dynamics.

**Folic Acid Conjugated DNA Icosahedron**

Folic acid (FA) is used as an endocytic ligand to facilitate targeting of the DNA icosahedron towards a specific cellular pathway and recently demonstrated experimentally by Bhatia *et al.*[8] In order to study the molecular display characteristics of chemically conjugated FA on the DNA icosahedron, we have realized an atomistic representation of the FA-conjugated DNA icosahedron ($I_{FA}$). The icosahedron structure was equilibrated for 50 ns prior to the attachment of folic acid. We conjugated, Gaussian optimized structure of deprotonated FA at seven different nucleotide positions along a single edge of the DNA icosahedron.[60] In order to compare closely with the experimental results, we choose the same edge (V4 oligo) and nucleotide residues of the icosahedral DNA to conjugate FA as investigated in the recent experimental study. These specific nucleotide residues were modified to display amine groups on which FA molecules were attached using the xLEaP module of AMBERTOOLS[25]. The structure was solvated using a truncated octahedral TIP3P water box. The details of the system building and simulation protocol are described in the appendix S8 of SI. We found that the dynamics of the DNA icosahedron is largely unaffected by the conjugation of FA ligands. Figure 8(a) shows a snapshot of $I_{FA}$ after 60 ns of MD simulation. Only FA molecules (shown in Figure S8(b) in SI) are highlighted in the vdW representation in Figure 8(c) with the same number tag corresponding to the experimental structure. We calculated the RMSD and RMSF for all seven FA positions with respect to the initial energy minimized structure to measure the deviation from the initial structure. Figure 8(d) shows the time evolution of RMSD for individual FA residues. The RMSD of the complete structure is shown in Figure S8(c) in the SI. We observe that stacking interaction stabilize the FA residue in the vicinity of DNA. Previous experimental studies show that stacking of FA can drive its self-assembly in vitro.[61] Here, FA residues stack between the adjacent DNA bases mediated by π-π stacking interactions. From the trajectories of the simulation, it is notable that in positions where the FA molecules are displayed facing the internal cavity of the DNA icosahedron (position 5 and 7) there are greater stacking interactions with the DNA bases. Further, geometrical analysis of the helical axis of the duplex DNA edge and the center of mass of the FA residues clearly indicate those FA residues at positions 3, 11 and 13 projects outwards (shown in cyan, blue and red in figure 8(c)). This analysis is in good agreement with experimental verification of FA-tag accessibility evidenced by cellular uptake efficiency. Figure 8(e) shows the RMSF with respect to the position index of FA over the entire MD simulation.

**Summary and Conclusions**

We have developed an atomistic model of an experimentally realized DNA icosahedron and investigated its solution behavior *in silico*. We presented 205 ns long simulations of the DNA icosahedron with explicit solvent and ions. The host-guest interaction of AuNPs entrapped within a DNA icosahedron has also been studied for the first time using atomistic simulations. We also follow the properties of endocytic tags, such as FA, displayed on the surface of the icosahedron. Our analysis of the simulation revealed that after some initial deviations from the built configuration, the nanostructures attain a robust and steady thermodynamic state. PCA analysis on simulation trajectories identified the major mode of motion in the structure and compares the relative fluctuations. Importantly, our studies revealed that AuNPs encapsulated within DNA icosahedra has relatively lower structural and dynamical fluctuations. The AuNPs are rigid bodies that restrict the motion of the icosahedral DNA shell which in turns reduces the dynamical fluctuations while compared to empty DNA icosahedra. The variation in the geometrical parameters of duplex DNA domains of the icosahedra in comparison to the B-DNA shows that the former maintain their B-DNA geometry quite well, showing only predictable conformational fluctuations arising due to the unpaired bases at the 5WJs or vertices. The analysis of the simulation trajectories backs up several experimental findings of the DNA icosahedron such as the end to end diagonal distance, cavity size etc. The diagonal distance analysis indicates that the structures elongate along one diagonal while compacting across the other. Our results confirm that the encapsulated AuNP remain intact and are structurally unaffected by the surrounding icosahedral DNA scaffold. Although, the experimental studies show the stability of 5WJ, the atomic level understanding is still not complete.[62, 63] Our model can be further used to investigate the symmetry effects and the junction stabilities. Surface chemistry of the nanoparticles and the nature of interactions (hydrophobic vs hydrophilic) play important role in the structure of DNA-nanoparticle composite.[64] It will require further studies to explore the exact nature of the interaction and stabilizing influences of the various encapsulated nanoparticles such as silver nanoparticle (AgNP) or silica nanoparticle (SiNP) on the structure of DNA icosahedra. A coarse-grained MD simulation study can help to better understand the self-assembly process of oligonucleotides into a complex DNA icosahedron.[19,65] The current computational study resolves key experimental observations such as (i) the reason why FA tags at specific nucleotide positions undergo better uptake than others (ii) why encapsulation of AuNPs within DNA half-icosahedra gave yields which were unexpectedly high that could not be explained. The atomistic model and dynamics of this system provides a structural map for researchers to display a variety of molecules on the surface and understand the nature and properties of cargo that can be encapsulated in this polyhedral capsule. This is important as, so far, this DNA icosahedron is only DNA polyhedron that is capable of simultaneously encapsulating cargo and displaying functionalities, that is also formed in quantitative yields and is structurally stable under physiological conditions.

**Acknowledgement**

We thank DAE, India for financial help. We acknowledge Supercomputer Education and Research Center (SERC) IISc Bangalore for providing access to high performance supercomputer SahasratT. H.J. thanks to CSIR, GoI for research fellowship. DB thanks CSIR, GoI for Research fellowship and French Embassy for Charpak fellowship. DB thanks Institute Curie and HFSP for long-term postdoctoral fellowships. We acknowledge Vishal Maingi, Atul Kaushik and Nilmani Singh for their help in the initial model building of the structures.

**Figures and Table**

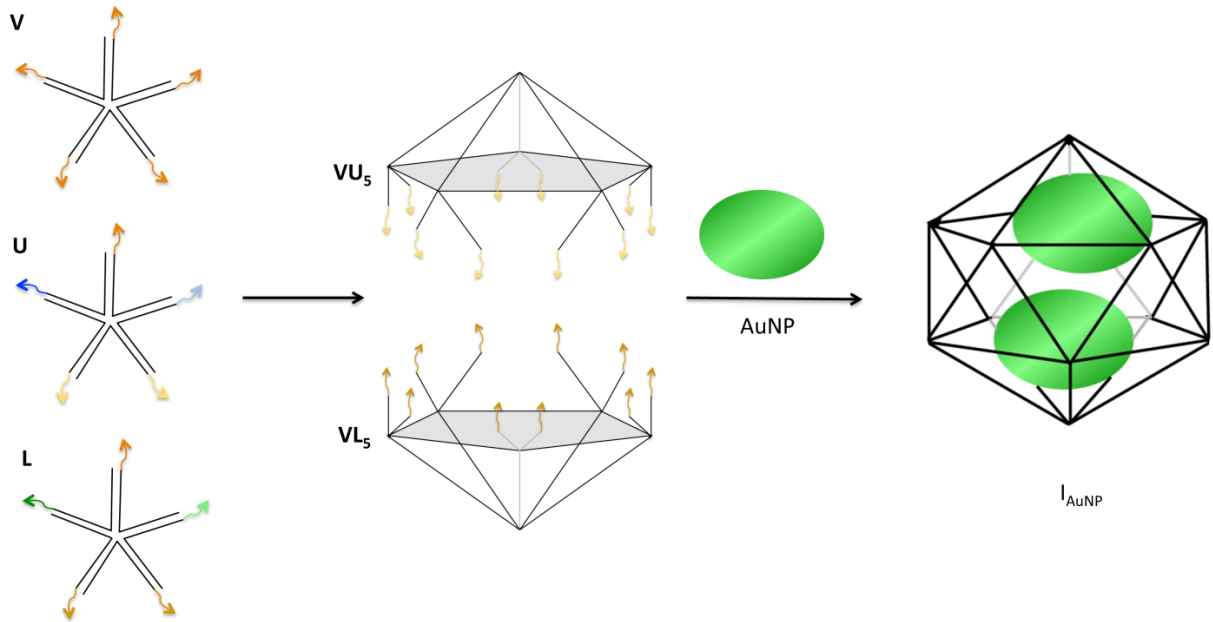

**Figure 1:** The Schematic Representation to the Building of DNA icosahedron:

The 5WJ self-assembles in 1:5 stoichiometry with 5WJs U and L (left) to yield respective half icosahedra called $VU_5$ and $VL_5$ (middle). The complimentary single strands are color-coded. The two half icosahedra VU5 and VL5 self-assemble in 1:1 stoichiometry to yield full icosahedron. External AuNPs can be encapsulated during this final assembly to yield AuNP encapsulated icosahedron ($I_{AuNP}$,) on the right.

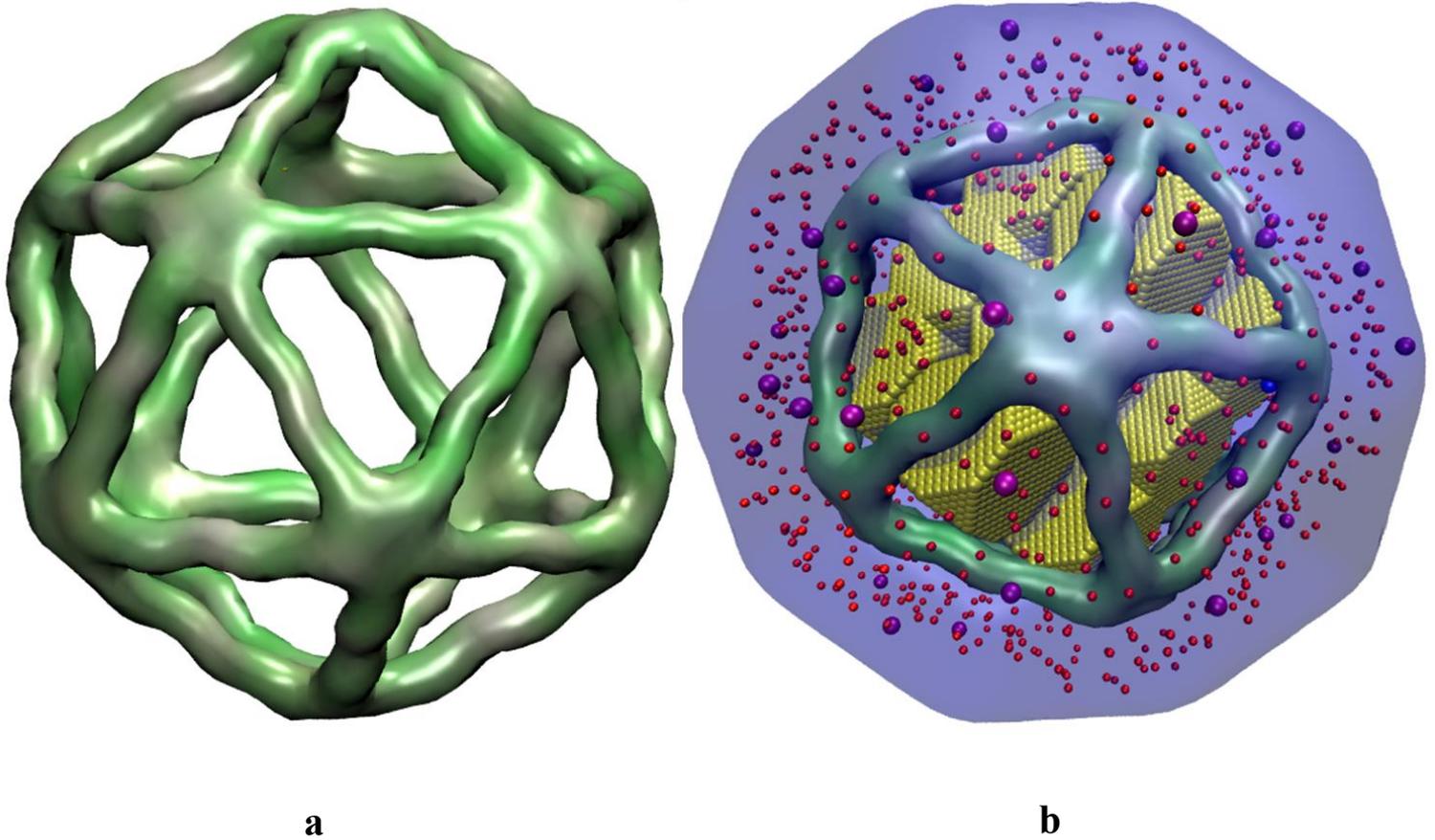

**Figure 2:** Initial Structure of the DNA icosahedron

(a) NAB built structure of empty icosahedral DNA ($I_{empty}$) (b) Au nanoparticle encapsulated icosahedral DNA ($I_{AuNP}$) with ions and water box. The surface representation with a probe radius of 5 Å, is used to highlight all the dSDNA edges. AuNPs and counter-ions have been shown in the van-der Walls (vdW) representation. $Mg^{2+}$ and $Cl^-$ ions are shown with larger vdW radii. The truncated octahedral water box is shown as continuous background. The simulated system corresponding to $I_{empty}$ contains ~0.78 million atoms including water and ions.

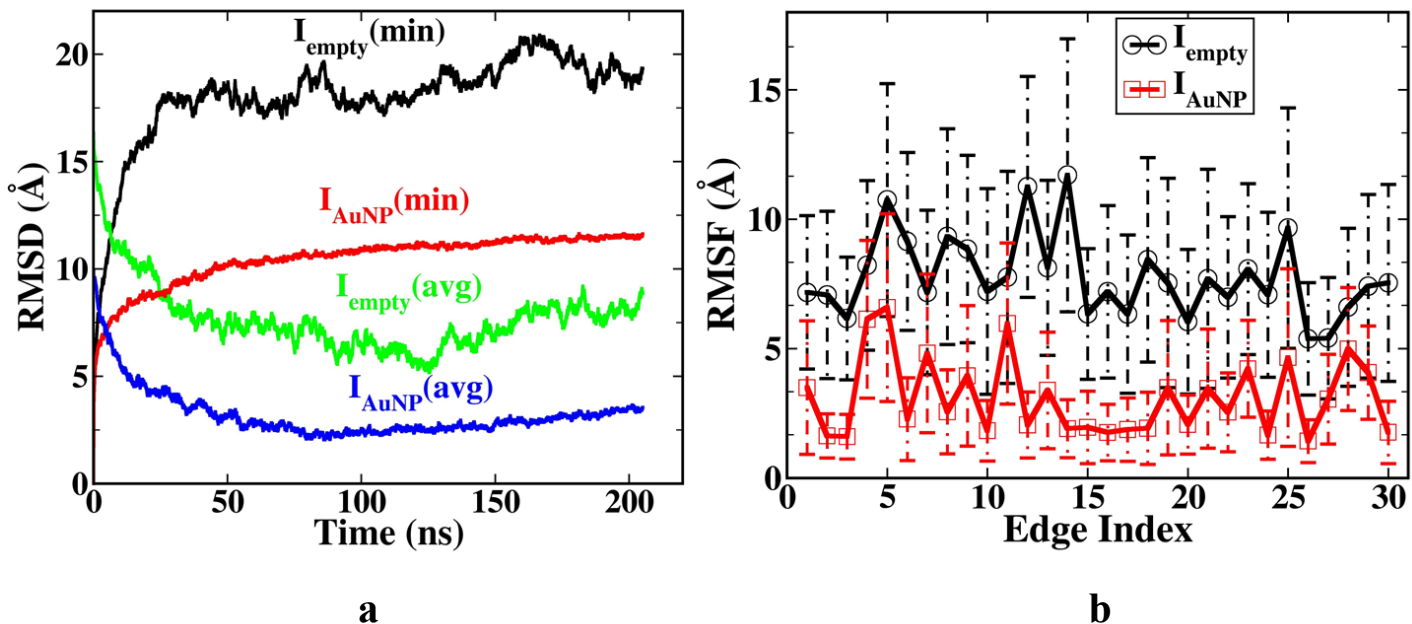

**Figure 3:** RMSD and RMSF of the Structures during the Simulation:

(a) The time evolution of the RMSD of icosahedral DNA with respect to the initially minimized structure and also with respect to the average structure over the last 50 ns long simulation. (b) RMSF: time average of the atomic positional fluctuation in the respective edges of icosahedron calculated over the whole simulation trajectories. $I_{AuNP}$ shows lesser structural deviations.

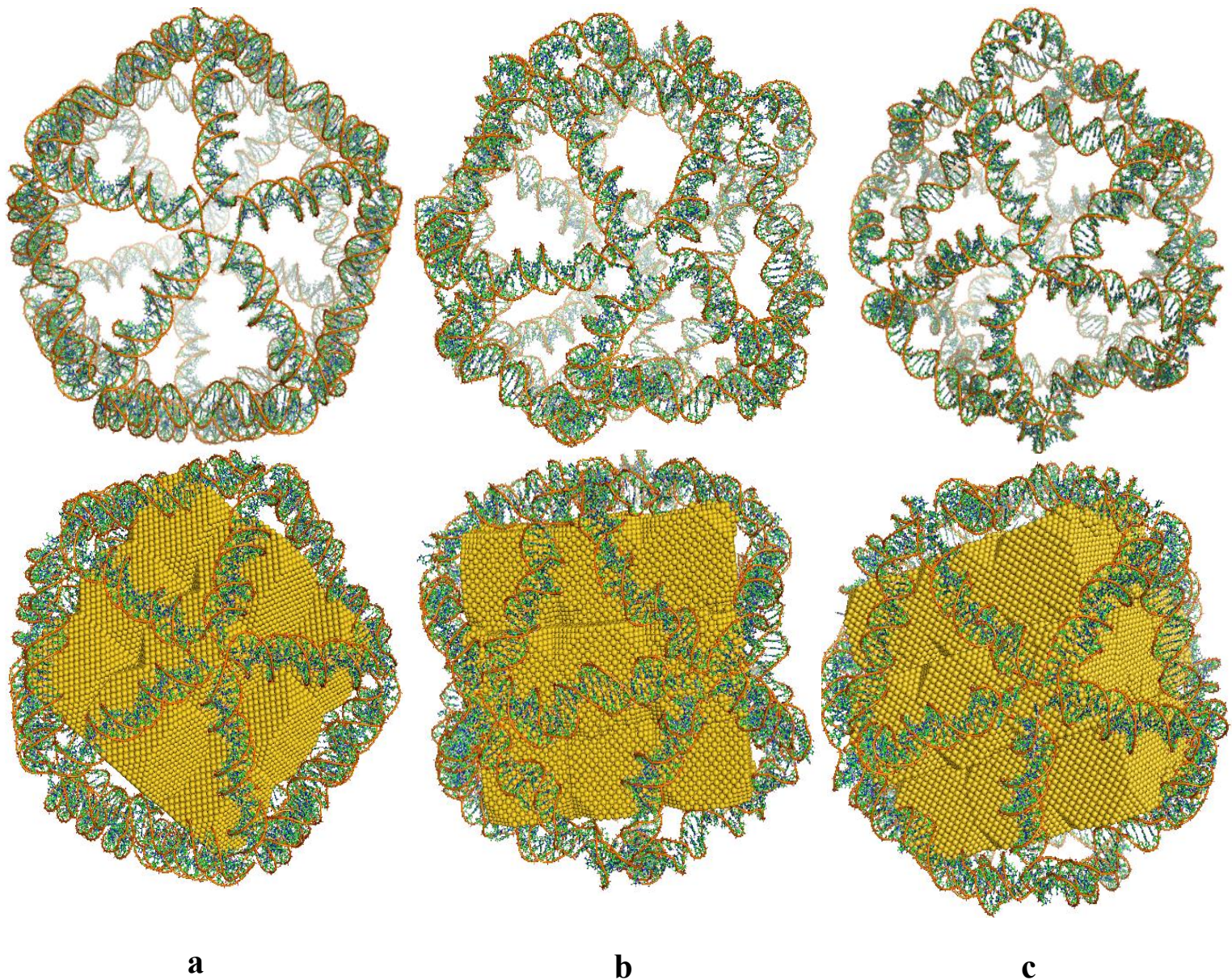

**Figure 4:** Instantaneous Snapshots of Structures during the Simulation:

(a) Energy minimized structure prior to MD simulation, (b) structure after 200 ns long MD simulation and (c) the structure averaged over whole the simulation trajectory. The top and bottom row correspond to $I_{empty}$ and $I_{AuNP}$ DNA respectively. DNA and gold atoms are shown in stick and bead representations respectively. For the sake of clarity, ions and water are not shown in the above snapshots.

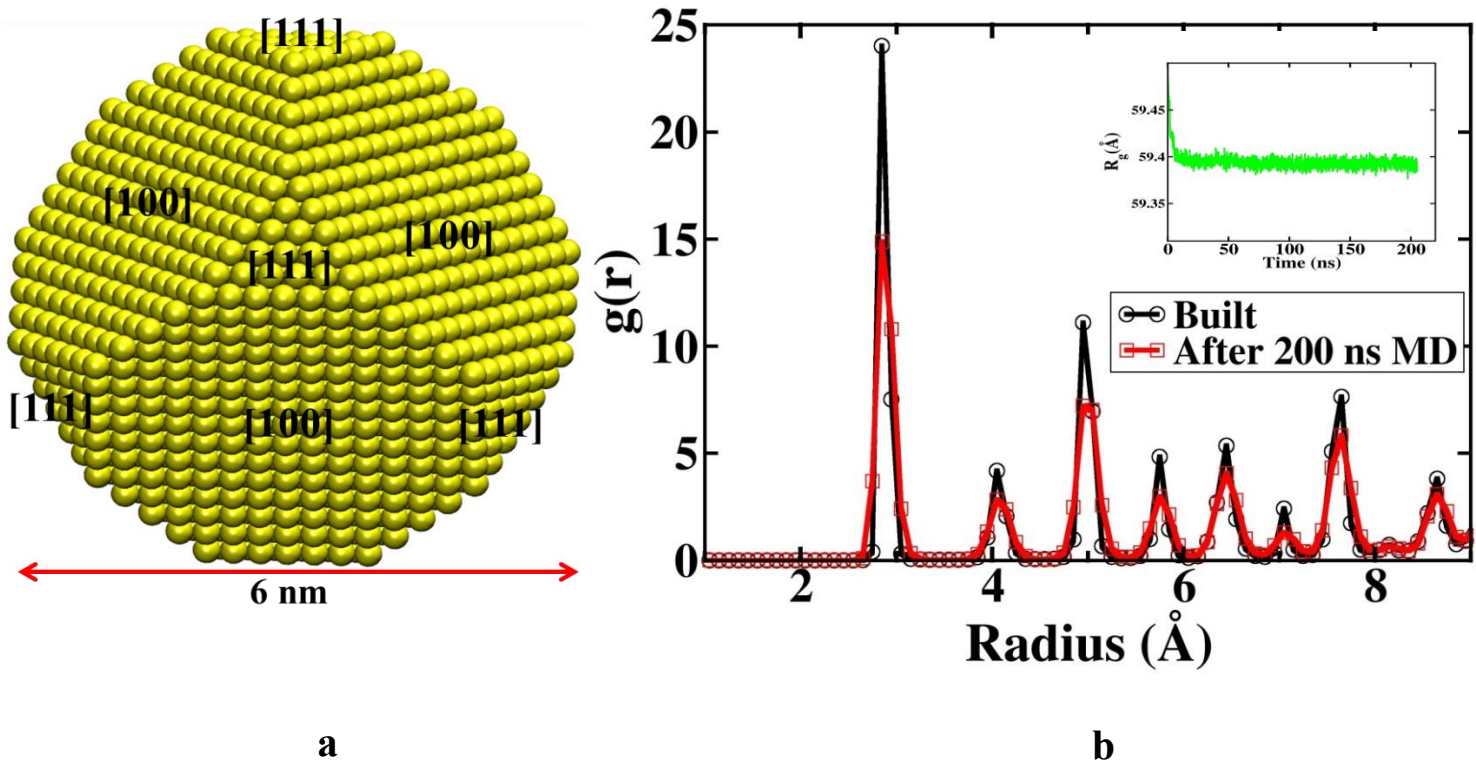

**Figure 5:** The Structure of AuNPs:

(a) The faceted gold nanoparticles (AuNP) generated using material studio 6.0. The AuNP icosahedral DNA has 21 such nanoparticles closely packed inside the icosahedron. (b)The radial distribution function, g(r) of AuNPs inside the icosahedral DNA compared between the built structure and after 205 ns long MD simulation. The radius of gyration ($R_g$) of all the nanoparticles inside the icosahedron along the simulation trajectory is shown on the upper right corner. All the nanoparticles maintain their crystallinity throughout the simulation.

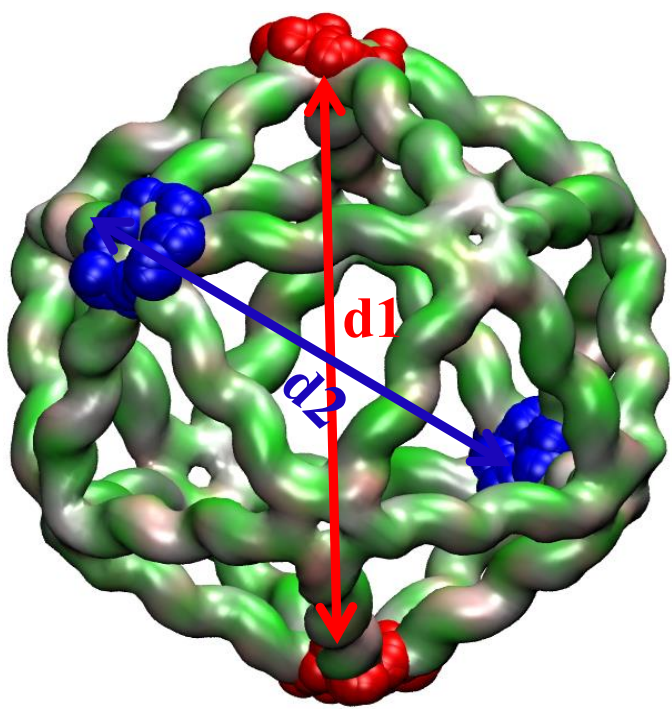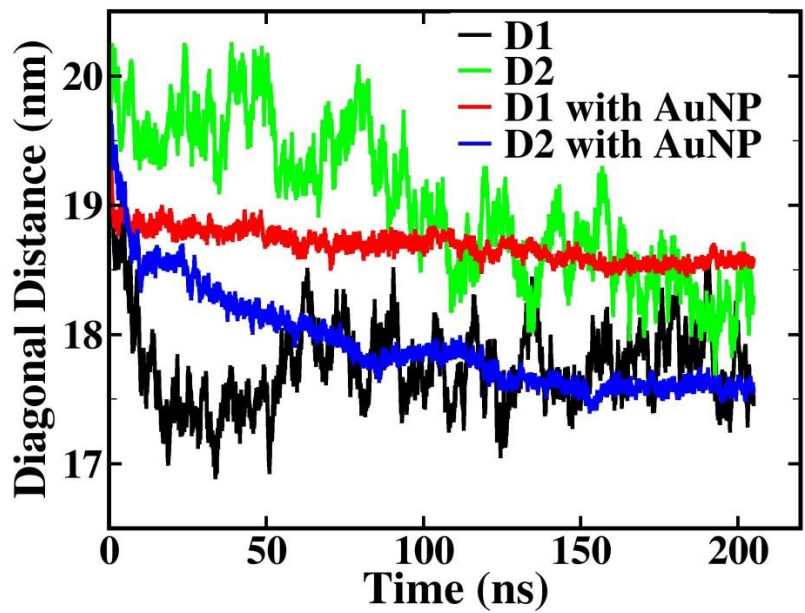

**a** **b**

**Figure 6:** The End to End Distances:

The diagonal distances of icosahedral DNA; first the geometrical centers of the all 5 terminal base pairs of vertices is calculated and the distance between geometrical centers of the furthest vertices is measured. (a) D1 and D2 are the respective diagonal distances between two different diagonals. The corresponding base pair residues are highlighted in red (D1) and blue (D2) representation in the figure whereas the edges of the icosahedron are shown in the surface representation. (using a probe radius of 3Å) (b) The time evolution of the diagonal lengths D1 and D2 in $I_{empty}$ and $I_{AuNP}$ has been shown along with the simulation time.

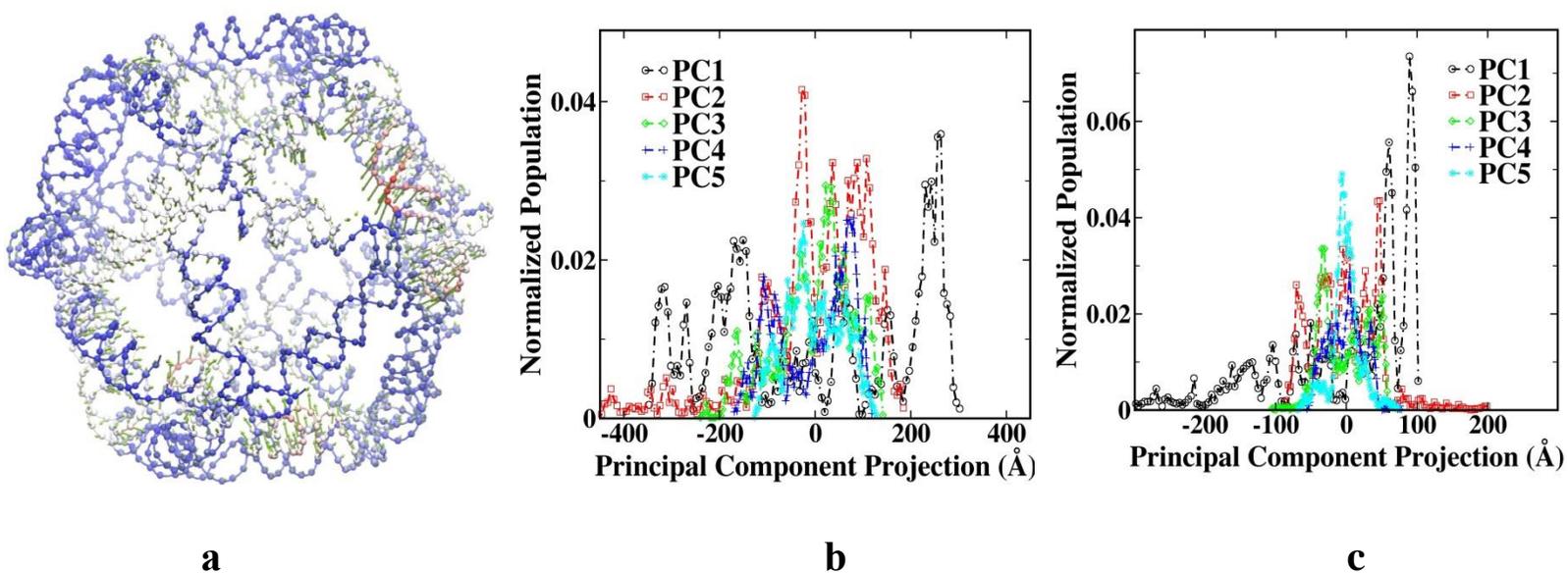

**Figure 7:** Principal Component Analysis

(a) Representation of the largest mode of the motion of DNA icosahedron (38% of total) extracted through the principle component analysis on the trajectories of the phosphorus atoms. Arrows indicate the first eigenvector while the phosphorus atoms have been shown by beads in normal mode wizard tool of VMD. Histogram for the projection of trajectories on first five principal components are shown in (b) for $I_{empty}$ and in (c) for $I_{AuNP}$. These 5 components accounts for ~65 % of the total motion (As shown in the Figure S6 in the SI). The lower projection of $I_{AuNP}$ is attributed to the lesser fluctuation in the system.

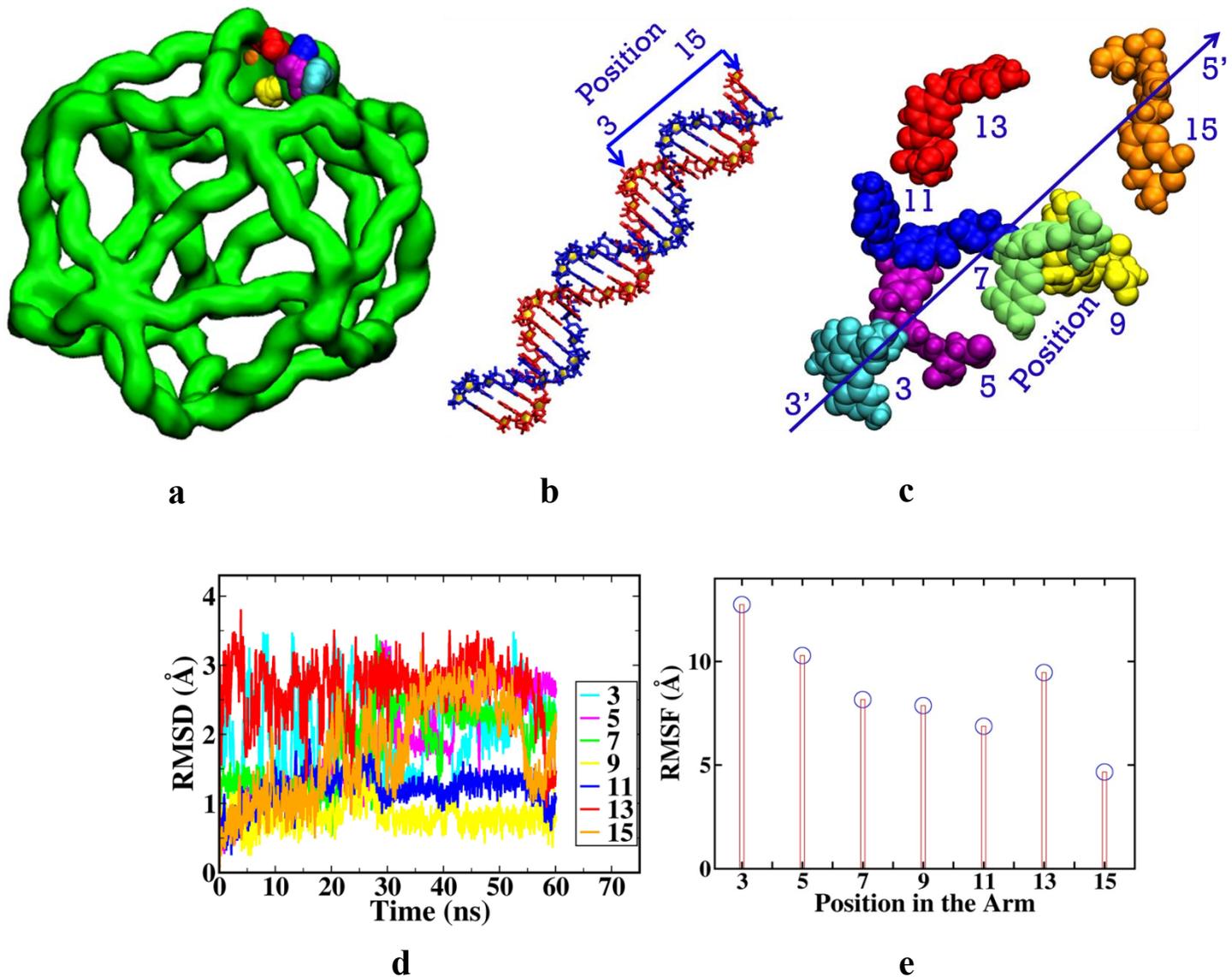

**Figure 8:** The Folic Acid Conjugated DNA Icosahedron.

(a) The FA conjugated DNA icosahedron ($I_{FA}$) after 60 ns MD simulation. FA molecules, covalently attached to 7 alternative nucleotides in one edge (V4), is shown in the vdW representation, whereas DNA has been shown in surface representation. (b) The edge where all FA molecules are attached to alternative positions. (c) The respective position of all FA residues after 60 ns MD simulation along the dSDNA edge as shown in Figure (b). (d) The time evolution of RMSD for all the seven FA residues in conjugation with DNA icosahedron during the course of MD simulation with respect to the initially minimized structures. (e) RMSF of individual FA ligands around their mean position during MD simulation. The snapshots have been fitted to the initially minimized structures to remove the rotational and translational motions before calculating RMSD and RMSF.

# Table

## Table 1

### Details of the Simulated Systems

|  | $I_{empty}$ | $I_{AuNP}$ |
|---|---|---|
| Total number of atoms | 780841 | 646078 |
| Number of atoms of DNA | 51964 (1632 bases) | 51964 (1632 bases) |
| Number of $Na^+$, $Mg2^+$, $Cl^-$ ions | 1632, 24, 48 | 1632, 24, 48 |
| Au atoms | - | 93555 (21 NP) |
| Water atoms | 727443 | 499125 |
| Truncated-octahedral box dimension | [224 224 224] Å [109.47° 109.47° 109.47°] | [218, 218, 218] Å [109.47° 109.47° 109.47°] |

**Supplementary Information**

**1. Coordinates and dimensions of the DNA icosahedron**

Each edge of the DNA icosahedron consists of a DNA helix of 26 base pairs and each vertex is made by joining five such helices with two unpaired bases at three positions at each vertex. We assume the vertex to be a circle with circumference equal to five times of the diameter of a DNA helix. The radius of this circle turns out 16 Å (by assuming the diameter of a B-DNA helix to be 20 Å)

Thus, the side length could be estimated to be 120.4 Å as follows,

(26*3.4) Å + (2*16) Å = 120.4 Å

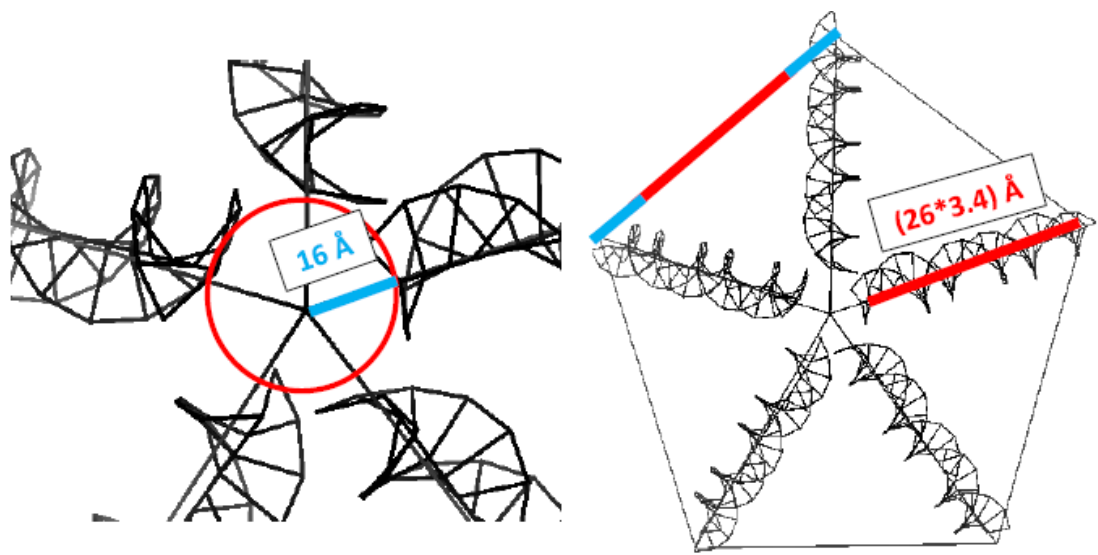

**Figure S1: Schematic showing the calculation of dimensions of the edges of icosahedron:**

The vertex of this 5WJ DNA polyhedral is assumed to be a circle of radius 16 Å. Thus the edge length of the ideal icosahedron has been approximated to 120.4 Å.. The representative images have been drawn using ChemBioDraw.

## 2. The vertices and full model of icosahedral DNA.

a

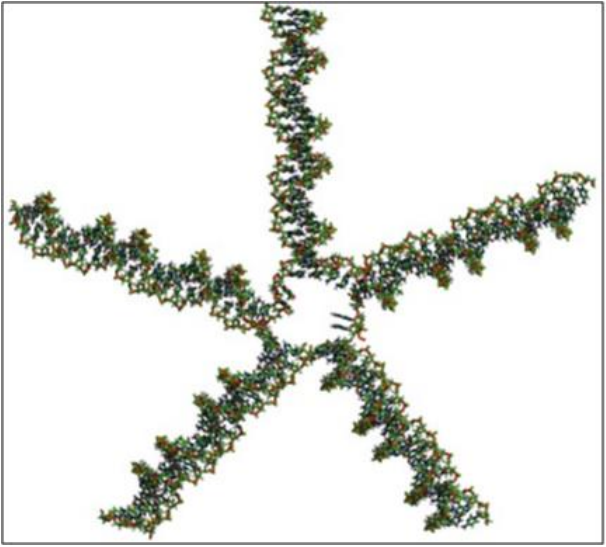 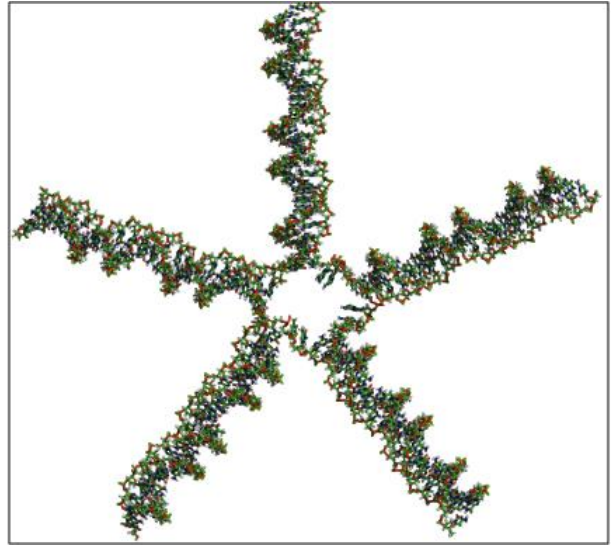

b

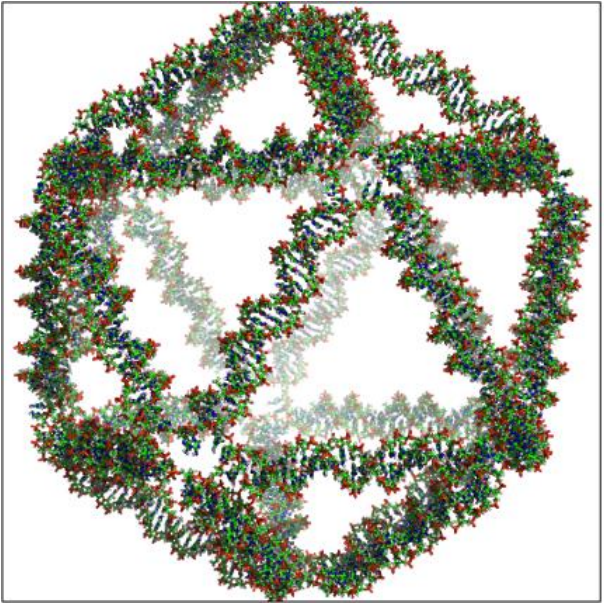 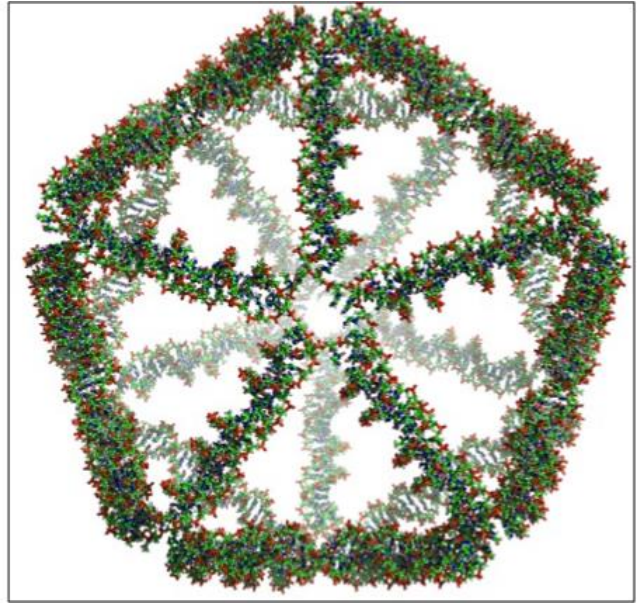

**Figure S2**

The atomistic model of a DNA 5WJ: (a) Top view (left) and bottom view (right) for one of the vertices. (b) Atomistic view of complete icosahedron showing C3 (left) and C5 (right) axes of symmetry.

## 3. The Build and Charge Neutralized Snapshots of Icosahedral DNA and AuNp Encapsulated Icosahedral DNA in Various Representations.

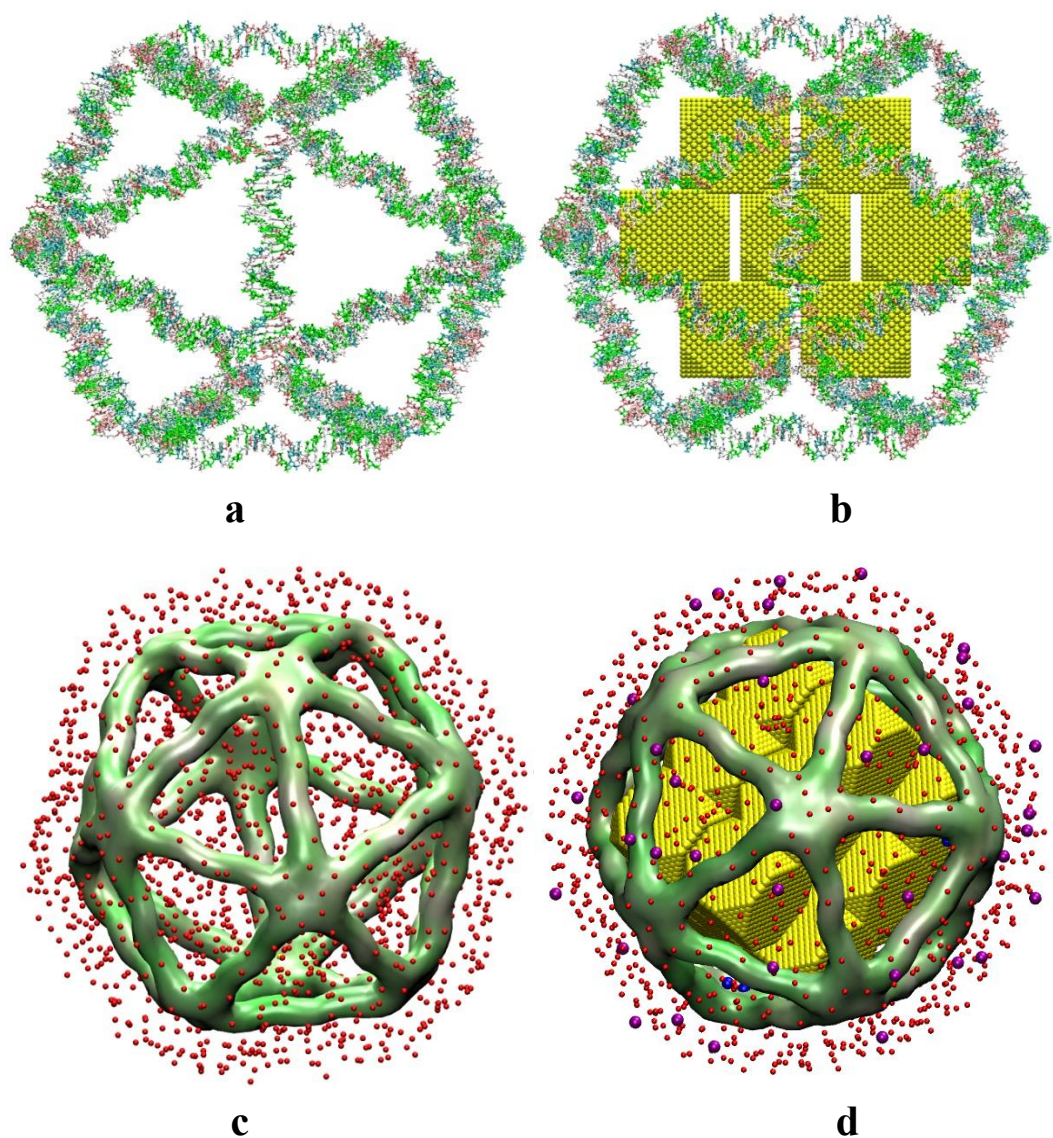

**Figure S3: Snapshots of the Built Structures in Different Representations**
NAB[1] built structure of (a) $I_{empty}$ and (b) $I_{AuNP}$. DNA has been shown in bond representation. The placement of ions around (c) $I_{empty}$ and (d) $I_{AuNP}$ structure is done using the xleap module of AMBER. The ion is placed by constructing a columbic potential grid and putting the ions at electrostatically favorable positions.

## 4. Instantaneous Snapshots of the Simulated Structures.

The snapshots of simulation at various instant of time of atomistic simulation have been shown in the figure S4. The structure of $I_{AuNP}$ is less distorted from its built configuration while compared $I_{empty}$

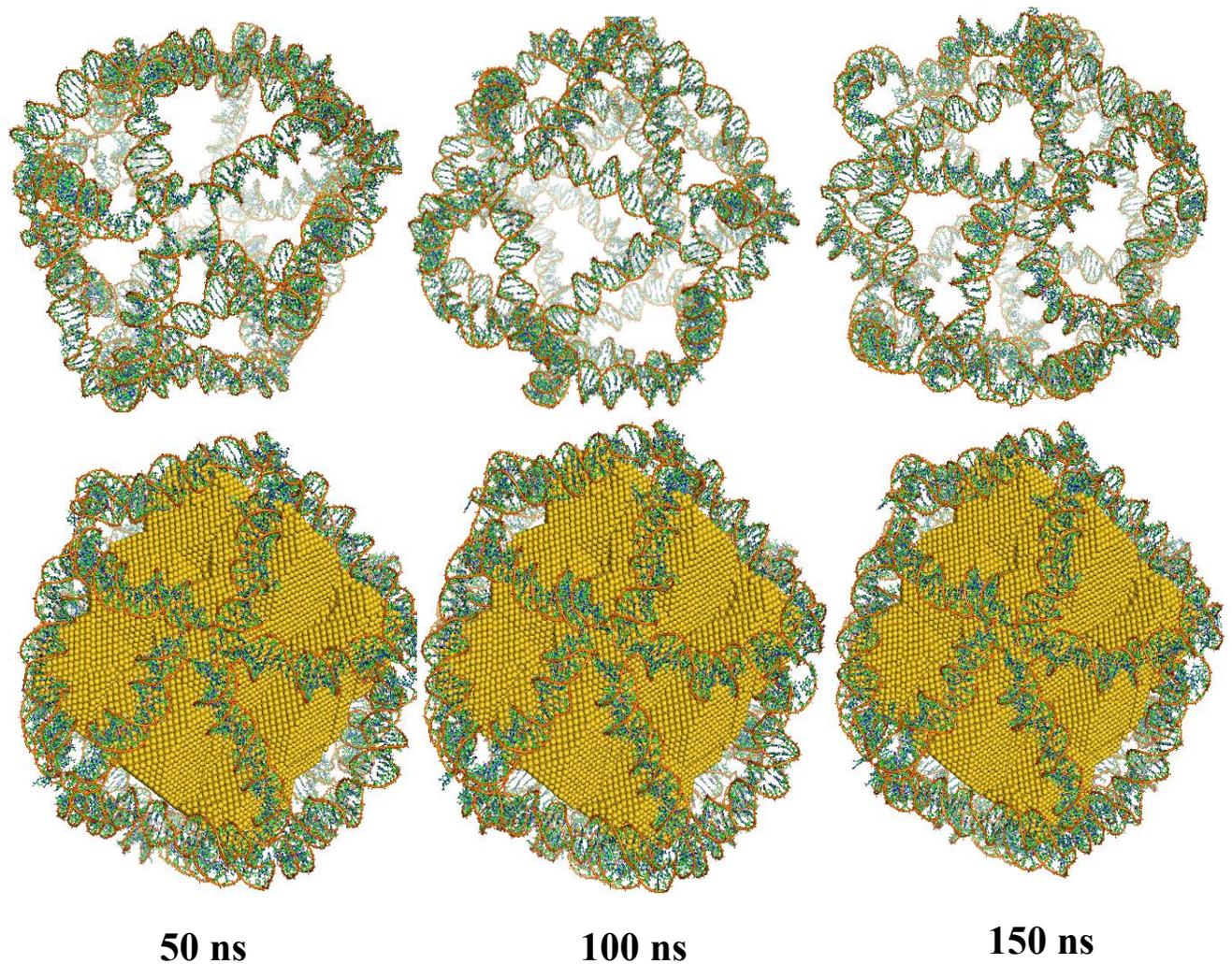

**50 ns**  **100 ns**  **150 ns**

**Figure S4: Structure of the DNA Icosahedron during the Simulation.**
The instantaneous snapshots of $I_{empty}$ and $I_{AuNp}$ after 50ns, 100 ns and 150 ns MD simulation are been shown in the top and bottom panel respectively.

## 5. Stability of Structures: RMSF *per-Resisdue* and RMSF *per-Atom*

Root-Mean-Square-Fluctuation (RMSF) values have been measured for all atoms of DNA icosahedron during the MD simulation. Figure S5 (a) and S5 (b) compares per residue and per atom RMSF for the all the residues and atoms of simulated structures respectively. The RMSF values for phosphorus atoms have been highlighted in figure S5 (b) with green and blue dots. The sinusoidal nature of the RMSF reflects various edges and junctions. The RMSF values of $I_{AuNP}$ are always lesser than $I_{empty}$

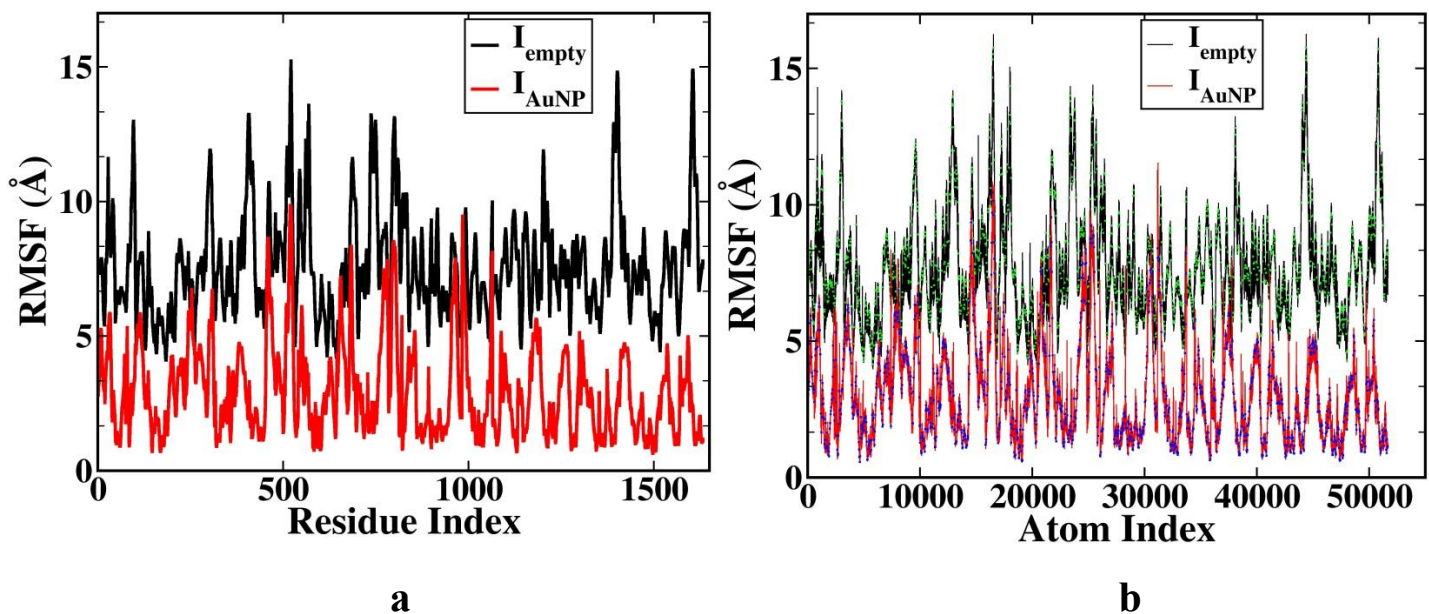

**Figure S5:**
(a) RMSF values per-residue (b) per atom RMSF averaged over 205 ns MD simulation for simulated structures. The RMSF values for phosphorus atoms have been highlighted in figure (b) with green and blue dots for $I_{empty}$ and $I_{AuNP}$ respectively. The RMSF values for $I_{AuNP}$ is lesser than $I_{empty}$ in both the cases.

## 6. Eigenvalues and Principle Component Analysis.

Principal component analysis has been performed on phosphorus atoms to identify the major modes of motion in dynamics of DNA nanostructure. It is observed from the analysis of the eigenvalues of the coordinate covariance matrix that first eigenvector accounts for the 38% and 47% of the total motion in case of $I_{empty}$ and $I_{AuNP}$ respectively. Since the sum of the eigenvalues is analogues of the total motion of the system, the fraction the eigenvalue to the trace (sum of eigenvalues) of matix gives the contribution to the motion. Figure 6 (a) and (b) shows first 50 eigenvalues out of the pool of 4896 eigenvalues and their contribution to the total motion respectively.

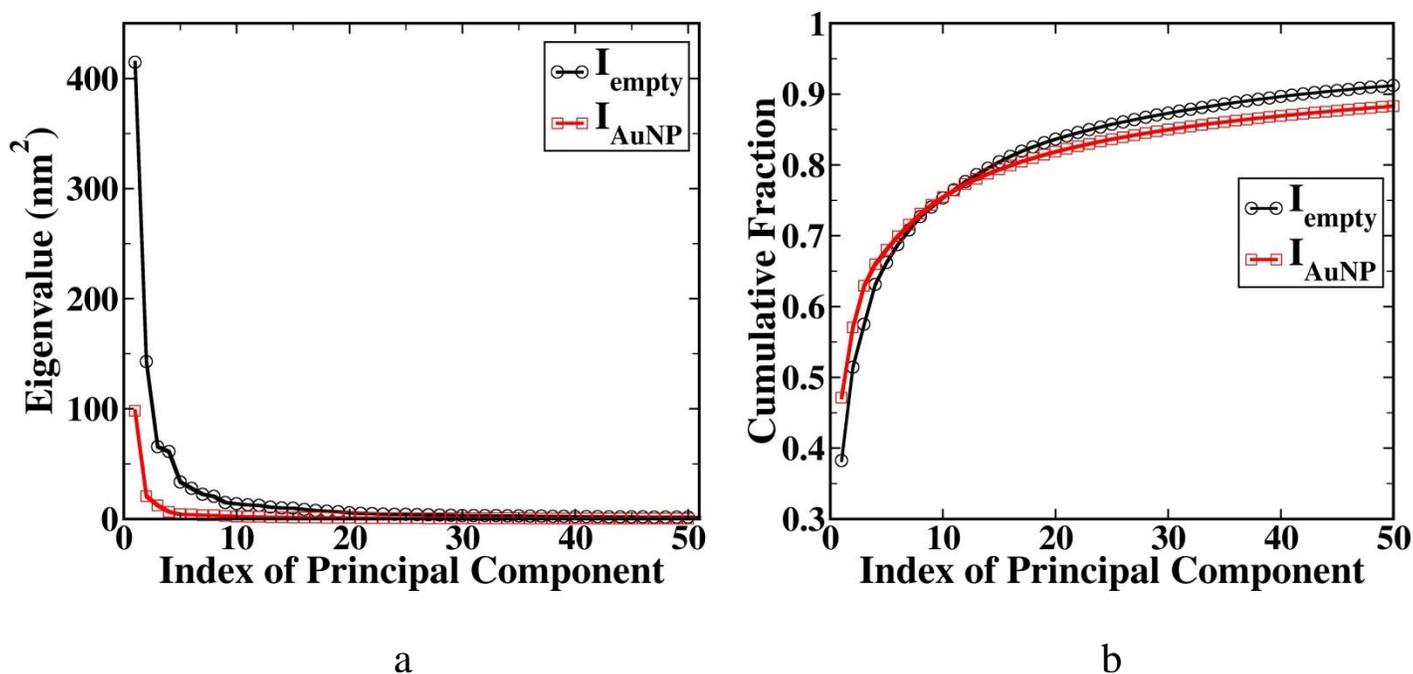

a
b

**Figure S6:**

(a) First 50 eigenvalues of the covariance matrix of Cartesian coordinates of phosphorus atoms during the whole simulation trajectory, (b) the cumulative contribution of corresponding eigenvectors to the motion has been shown. It is evident that first 50 eigenvectors (out of 4896) contributes for almost 90% of the total motion.

7. Time evolution of Radius of Gyration and Solvent Accessible Surface Area.

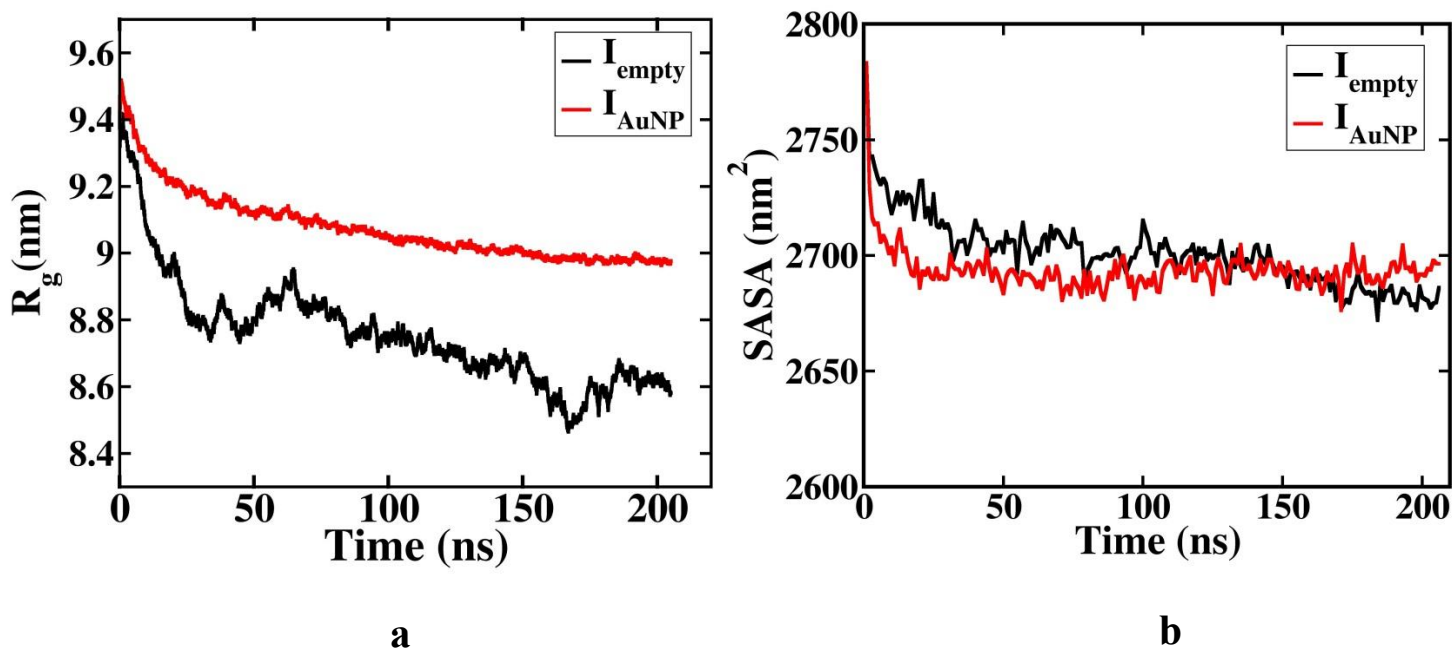

**Figure S7: Time Evolution of $R_g$ and SASA:**

(a)Time evolution of the radius of gyration, $R_g$, of the DNA icosahedron. (b) Solvent accessible surface area, SASA, of all the atoms of icosahedral DNA using LCPO method with probe radius 1.4 Å. The analysis shows that the structures maintain icosahedral geometry during the course of simulation and also highlights the cargo bearing capability of the DNA icosahedron.

## 8. Conjugation of Folic Acid to DNA Icosahedron.

Folic acid (FA) has been used as a biological tag to study the endocytic pathways of cargo loaded DNA icosahedron. In order to probe the microscopic details of folate ligands to DNA icosahedron, we have come up with the atomistic model of the structure. First, we optimize the deprotonated pdb structure of FA using Gaussian09[2]. Here, we find the wave-function of atomic orbitals via Hartree-Fock first principle calculations using 6/31G basis set to get the optimized structure (shown in figure S8 b). The partial atomic charges for the atoms of folate residue were calculated using restrained electrostatic potential (RESP) approach[3]. We have used the GAFF[4] parameters to describe bonded and non-bonded interactions among FA atoms. These parameters have been generated using the antichamber module in AMBER programming suit[5]. Following the standard MD protocol, as mentioned in the simulation methodology section, we minimize and then equilibrate the FA molecule through 50 ns production run. We took the structure of the folate residue after this MD simulation to conjugate into DNA icosahedron. Prior to this conjugation, we equilibrate the structures of DNA icosahedron for 50 ns with explicit ions and water. We slightly modify the bases of DNA at some sites with amine group to attach the FA residues to icosahedron in xleap[5]. We chose the 7 sites to attach the FA residues into one particular edge of DNA icosahedron to compare the experimental results from Bhatia et al.(Bhatia et al.,Nat. Nanotech. in press) The system is charge neutralized with Na+ ions and additionally 12 Mg2+ and 24 Cl- ions are added for the junction stabilities. We solvate the charge neutralized DNA icosahedron into an octahedral TIP3P water box ensuring 10 Å solvent shell around solute. We perform a series of energy minimization steps to remove the bad contacts in the system. The system is gradually heated up to 300 K temperature with weak harmonic restraints (20 kcal/mol/Å$^2$) to DNA and folate residues. Further, the system was equilibrated using NPT ensemble to attain the desired density. This is followed by 60 ns production run with no harmonic restraints to the system. We use 1 fs time step to integrate the equation of motion with 1ps$^{-1}$ time constant for temperature coupling with heat bath in Berendson thermostat. We use particle mess Ewald molecular dynamics (PMEMD) approach integrated with AMBER programming environment to calculate the long range electrostatic interaction in MD simulation with periodic boundary conditions. Figure S8 (a) shows the snapshot of the structure at the end of the simulation, the folate residues has been highlighted in sphere representation while the DNA strands are shown in the surface representation. The structure of folic acid has shown in the figure S8 (b). Figure S8 (c) shows the time evolution of RMSD with respect to initially energy minimized structure. We observed the dynamics and interaction of FA molecules in conjugation to DNA icosahedron. After some deviation from initial structure, the folate conjugated DNA icosahedron ($I_{FA}$) takes a thermodynamically stable configuration.

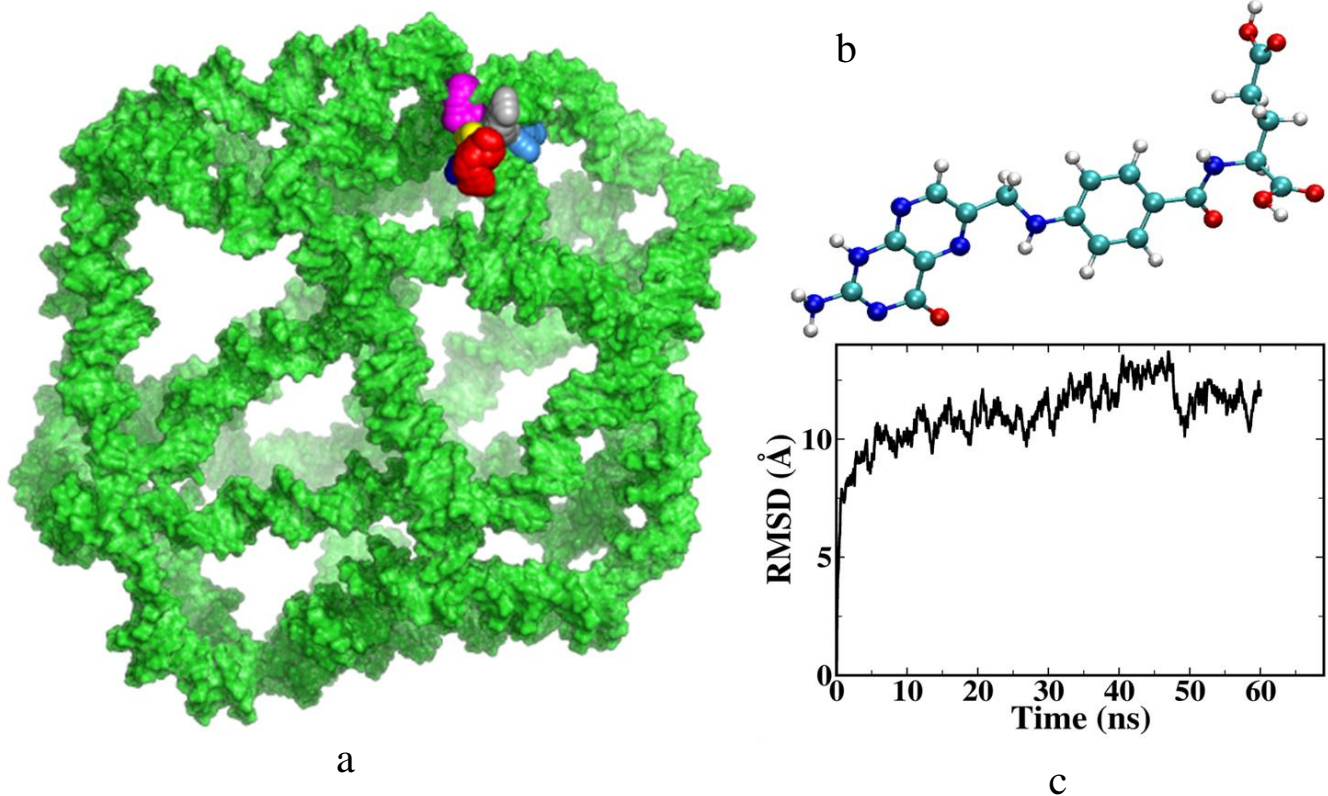

**Figure S8**

(a) Structure of FA conjugated DNA icosahedra after 60 ns MD simulation, the folate residues are highlighted in sphere representation and DNA strands have been shown in green surface representation. (b) Atomistic representation of folate residue. (c) The RMSD evolution of FA conjugated DNA icosahedron.

## 9. Conformational Analysis: Average Geometrical Parameters for the Edges of Icosahedral DNA.

We calculate various DNA helical parameters for all the duplex DNA domains comprising the edges of the DNA icosahedron in order to estimate how well the helical geometry is preserved during the simulations. The overall fluctuations in the global structural parameters of DNA (base pair, base step and helicity) during the span of the MD simulation were quantified using CPPTRAJ, analysis utility in AMBER MD programming suite.[5, 6] Table S1a-c provided in appendix 9 of SI shows the base pair, base step and helical parameters for the duplex DNA domains in the DNA icosahedron averaged over all 30 edges respectively. The table compares the values of these three parameters at three different stages of simulations namely, prior to energy minimization, after energy minimization and after 205 ns MD simulations. The bottom row in each table shows the values of the respective parameters for 12-mer B-DNA built ($B_{built}$) using a standard NAB routine. Despite the relatively high standard deviation, the values of these parameters are comparable to the respective values for $B_{built}$. Also, all the parameters differ slightly from the structure pre-energy minimization to the structure after 205 ns MD simulation. This analysis essentially reveals that the geometry of the helix is well preserved during the simulations. A comparison of $I_{empty}$ and $I_{AuNP}$ revealed that the later structure shows less deformation from ideal B-DNA. We posit that the underlying extra stability of $I_{AuNP}$ arises from host-cargo interaction that reduces thermal fluctuations of the neighboring DNA atoms. The values of twist parameter (both base-step and helical twist) reduce while compared to the built structures which is manifested to the bending of DNA helices. We observe large values of standard deviation in the above parameters but the average values are in good agreement to respective reference B-DNA parameters. The standard deviation in the parameters is largely manifested in the terminal base-pairs i.e. the base pairs at the vertices of the icosahedron. This is expected since at the vertex, the strands in a given duplex DNA domain change their helical domain/axis. The unpaired bases at the vertices (excluded for this analysis) also contribute to the deformation of ideal parameters originating from the unwinding of terminal base-pairs.[7] The opposite ends of the edges offset the values of the parameter, so the average value is not very different with respect to the reference structures. While, above analysis has been averaged over all the edges of the DNA icosahedron, table S2a-c in the appendix 8 of SI gives the values of these parameters for an edge A1 (chosen randomly out of 30 edges). The analysis shows that apart from the fluctuation at the vertices, the geometry of DNA is better maintained in $I_{AuNP}$ during the course of simulation.

**TableS1: Geometrical parameters of DNA averaged over all the edges.**

**Table S1 (a)**

**Base-Pair Parameters for DNA Averaged Over all the Edges of DNA Icosahedron**

| Snapshots Time | Name of the Structure | Shear (Å) | Stretch (Å) | Stagger (Å) | Buckle (°) | Propeller (°) | Opening (°) |
|---|---|---|---|---|---|---|---|
| Built | Icosahedron | 0.04 (±0.45) | 0.21 (±0.33) | 0.36 (±0.82) | 0.64 (±12.26) | -5.37 (±8.80) | 4.19 (±15.91) |
| After energy minimization | $I_{empty}$ | 0.00 (±0.54) | 0.05 (±0.36) | 0.26 (±0.64) | 1.20 (±16.51) | -9.96 (±11.42) | -1.74 (±12.60) |
| | $I_{AuNP}$ | 0.00 (±0.63) | 0.06 (±0.44) | 0.33 (±0.75) | 0.90 (±18.63) | -11.44 (±12.32) | -0.72 (±13.54) |
| After 205 ns MD | $I_{empty}$ | 0.02 (±1.28) | -0.10 (±1.87) | 0.03 (±1.35) | -0.31 (±19.32) | -10.25 (±15.72) | 0.64 (±18.89) |
| | $I_{AuNP}$ | 0.01 (±1.62) | -0.08 (±1.12) | 0.08 (±1.30) | 0.51 (±21.50) | -9.95 (±18.90) | 2.39 (±24.20) |
| Pictorial Representation | 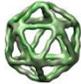 | 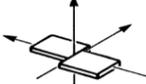 | 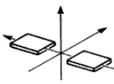 | 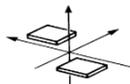 | 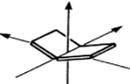 | 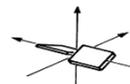 | 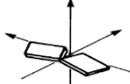 |
| Built | 12-mer B-DNA | 0.00 (±0.08) | 0.12 (±0.03) | 0.0 (±0.0) | 0.0 (±0.05) | 0.03 (±0.01) | 0.0 (±0.08) |

## Table S1 (b)
## Base-Step Parameters for DNA Averaged Over all the Edges of DNA Icosahedron

| Snapshots Time | Name of the Structure | Shift (Å) | Slide (Å) | Rise (Å) | Tilt (°) | Roll (°) | Twist (°) |
|---|---|---|---|---|---|---|---|
| Built | Icosahedron | 0.02 (±0.37) | -0.16 (±0.37) | 3.23 (±0.32) | 0.45 (±5.57) | -2.78 (±6.00) | 35.86 (±4.79) |
| After energy minimization | $I_{empty}$ | 0.01 (±0.64) | -0.26 (±0.63) | 3.25 (±0.27) | 0.00 (±5.59) | -1.87 (±5.56) | 35.96 (±6.34) |
|  | $I_{AuNP}$ | 0.02 (±0.74) | -0.22 (±0.67) | 3.26 (±0.37) | -0.04 (±6.51) | -1.78 (±6.73) | 35.85 (±6.69) |
| After 205 ns MD | $I_{empty}$ | -0.03 (±0.91) | -0.31 (±0.82) | 3.37 (±0.50) | 0.06 (±6.99) | 3.02 (±11.18) | 33.49 (±9.37) |
|  | $I_{AuNP}$ | 0.05 (±1.27) | 0.10 (±1.23) | 3.42 (±0.72) | -0.49 (±9.95) | 2.12 (±12.38) | 34.45 (±12.41) |
| Pictorial Representation | 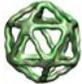 | 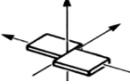 | 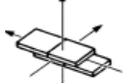 | 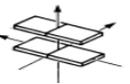 | 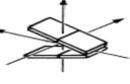 | 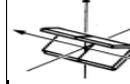 | 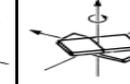 |
| Built | 12-mer B-DNA | 0.00 (±0.01) | -0.27 (±0.03) | 3.37 (±0.01) | 0.03 (±0.27) | 3.20 (±0.21) | 35.71 (±1.28) |

## Table S1 (c)
## Helical Parameters for DNA Averaged over all the Edges of DNA Icosahedron

| Snapshots Time | Name of the Structure | X-disp. (Å) | Y-disp. (Å) | Helical Rise (Å) | Inclination (°) | Tip (°) | Helical Twist (°) |
|---|---|---|---|---|---|---|---|
| Built | Icosahedron | 0.08 (±0.92) | 0.01 (±0.94) | 3.17 (±0.34) | -4.47 (±9.53) | -0.70 (±0.04) | 36.82 (±4.94) |
| After energy minimization | $I_{empty}$ | -0.21 (±1.16) | 0.01 (±1.37) | 3.20 (±0.35) | -2.80 (±9.11) | -0.29 (±0.04) | 36.80 (±6.38) |
| | $I_{AuNP}$ | -0.19 (±1.31) | 0.00 (±1.53) | 3.18 (±0.46) | -2.59 (±10.72) | -0.18 (±0.05) | 37.02 (±6.74) |
| After 205 ns MD | $I_{empty}$ | -1.16 (±1.95) | 0.00 (±1.86) | 3.22 (±0.50) | 5.96 (±14.68) | -0.22 (±0.06) | 34.91 (±12.68) |
| | $I_{AuNP}$ | -0.47 (±2.56) | -0.12 (±2.30) | 3.22 (±0.64) | 4.49 (±17.56) | 0.49 (±0.07) | 36.73 (±15.24) |
| Pictorial Representation | 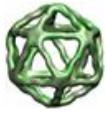 | 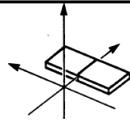 | 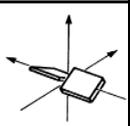 | 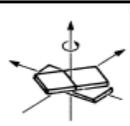 | 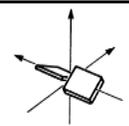 | 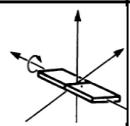 | 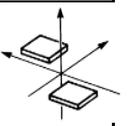 |
| Built | 12-mer B-DNA | 0.02 (±0.03) | 0.00 (±0.01) | 3.38 (±0.01) | -5.21 (±0.39) | 0.05 (±0.05) | 35.85 (±1.27) |

**TableS2: Geometrical parameters for an edge of DNA icosahedron**

**Table S2 (a)**

**Average Base-Pair Parameters for an Edge of DNA Icosahedron**

| Snapshots Time | Name of the Structure | Shear (Å) | Stretch (Å) | Stagger (Å) | Buckle (°) | Propeller (°) | Opening (°) |
|---|---|---|---|---|---|---|---|
| Built | Icosahedron | 0.00 (±0.35) | 0.27 (±0.30) | 0.46 (±0.77) | 1.08 (±8.66) | -3.72 (±5.86) | 9.94 (±15.76) |
| After energy minimization | I$_{empty}$ | 0.03 (±0.31) | 0.09 (±0.31) | 0.38 (±0.35) | 0.50 (±14.88) | -7.11 (±6.79) | 2.88 (±7.92) |
| | I$_{AuNP}$ | -0.02 (±0.52) | 0.09 (±0.35) | 0.47 (±0.65) | -0.12 (±15.77) | -9.32 (±11.25) | 3.67 (±9.77) |
| After 205 ns MD | I$_{empty}$ | -0.03 (±0.30) | 0.00 (±0.09) | 0.14 (±0.38) | -1.16 (±12.14) | -9.79 (±8.62) | -0.41 (±5.38) |
| | I$_{AuNP}$ | 0.20 (±0.74) | 0.00 (±0.14) | 0.17 (±0.26) | -1.18 (±11.31) | -8.97 (±18.21) | 2.70 (±9.64) |
| Pictorial Representation | 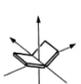 | 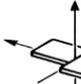 | 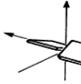 | 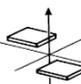 | 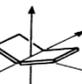 | 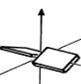 | 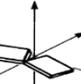 |
| Built | 12-mer B-DNA | 0.00 (±0.08) | 0.12 (±0.03) | 0.0 (±0.0) | 0.0 (±0.05) | 0.03 (±0.01) | 0.0 (±0.08) |

## Table S2 (b)

## Average Base-Step Parameters for an Edge of DNA Icosahedron

| Snapshots Time | Name of the Structure | Shift (Å) | Slide (Å) | Rise (Å) | Tilt (°) | Roll (°) | Twist (°) |
|---|---|---|---|---|---|---|---|
| Built | Icosahedron | -0.03 (±0.21) | -0.27 (±0.28) | 3.30 (±0.25) | 0.48 (±3.89) | -2.18 (±5.59) | 35.67 (±3.00) |
| After energy minimization | I$_{empty}$ | -0.05 (±0.43) | -0.38 (±0.42) | 3.24 (±0.21) | 0.39 (±3.57) | -0.85 (±4.32) | 36.66 (±3.36) |
| | I$_{AuNP}$ | -0.01 (±0.82) | -0.35 (±0.42) | 3.23 (±0.24) | 0.70 (±4.92) | -0.82 (±5.78) | 36.93 (±4.45) |
| After 205 ns MD | I$_{empty}$ | -0.02 (±1.06) | -0.14 (±0.65) | 3.27 (±0.30) | 0.63 (±4.08) | 1.54 (±6.92) | 34.46 (±4.76) |
| | I$_{AuNP}$ | 0.02 (±0.73) | -0.21 (±0.81) | 3.31 (±0.24) | -0.92 (±4.57) | 2.10 (±7.76) | 35.67 (±6.34) |
| Pictorial Representation | 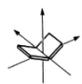 | 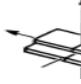 | 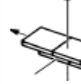 | 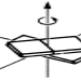 | 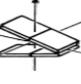 | 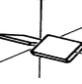 | 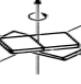 |
| Built | 12-mer B-DNA | 0.00 (±0.01) | -0.27 (±0.03) | 3.37 (±0.01) | 0.03 (±0.27) | 3.20 (±0.21) | 35.71 (±1.28) |

## Table S2 (c)

## Average Helical Parameters for an Edge of Icosahedral DNA

| Snapshots Time | Name of the Structure | X-disp. (Å) | Y-disp. (Å) | Helical Rise (Å) | Inclination (°) | Tip (°) | Helical Twist (°) |
|---|---|---|---|---|---|---|---|
| Built | Icosahedron | -0.10 (±0.86) | 0.12 (±0.70) | 3.25 (±0.24) | -3.78 (±9.00) | -0.73 (±0.19) | 36.37 (±2.86) |
| After energy minimization | $I_{empty}$ | -0.49 (±0.58) | 0.17 (±0.94) | 3.23 (±0.25) | -1.39 (±6.62) | -0.69 (±0.16) | 37.08 (±3.43) |
| | $I_{AuNP}$ | -0.46 (±0.90) | 0.15 (±1.54) | 3.17 (±0.39) | -1.14 (±9.40) | -1.11 (±0.20) | 37.72 (±4.16) |
| After 205 ns MD | $I_{empty}$ | -0.46 (±1.58) | 0.05 (±1.71) | 3.21 (±0.45) | 2.69 (±11.70) | -0.95 (±0.27) | 35.41 (±4.76) |
| | $I_{AuNP}$ | -0.84 (±1.81) | -0.25 (±1.24) | 3.18 (±0.38) | 4.49 (±12.87) | 1.50 (±0.29) | 36.87 (±6.31) |
| Pictorial Representation | 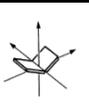 | 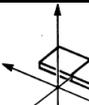 | 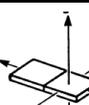 | 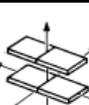 | 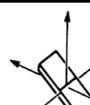 | 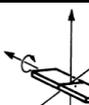 | 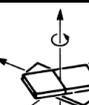 |
| Built | 12-mer B-DNA | 0.02 (±0.03) | 0.00 (±0.01) | 3.38 (±0.01) | -5.21 (±0.39) | 0.05 (±0.05) | 35.85 (±1.27) |